\documentclass[preprint,12pt]{elsarticle}

\usepackage{amssymb}
\usepackage{amsmath}

\journal{Solid State Communications}

\begin{document}

\begin{frontmatter}

\title{Anomalous \( 140 \, \text{K} \) Electronic Transition in Bi\textsubscript{2}Se\textsubscript{3}: Possible Charge Order in a Defect-Engineered System}

\author{Yanan Li$^{a,d}$\footnote{Email: liyanan@lntu.edu.cn. This paper is published as: Li, Y., Parsons, C., Ramakrishna, S. K., Dwivedi, A. P., Schofield, M. A., Reyes, A. R., \& Guptasarma, P. (2025). Anomalous 140 K electronic transition in Bi2Se3: Possible charge order in a defect-engineered system. Solid State Communications, 406, 2025,116176. https://doi.org/10.1016/j.ssc.2025.116176}, Christian Parsons$^{a}$, Sanath Kumar Ramakrishna$^{b,c}$, Anand Prashant Dwivedi$^{a}$, Marvin A. Schofield$^{a}$, Arneil P. Reyes$^{b,c}$, Prasenjit Guptasarma$^{a}$}

\affiliation[1]{organization={Department of Physics, University of Wisconsin-Milwaukee},
            city={Milwaukee},
            postcode={53211},
            state={Wisconsin},
            country={USA}}

\affiliation[2]{organization={National High Magnetic Field Laboratory},
             city={Tallahassee},
             postcode={32310},
             state={Florida},
             country={USA}}

\affiliation[3]{organization={Florida State University},
             city={Tallahassee},
             postcode={32310},
             state={Florida},
             country={USA}}

\affiliation[4]{organization={College of Electronic and Information Engineering, Liaoning Technical University},
            city={Huludao},
            postcode={125105},
            state={Liaoning},
            country={China}}

\begin{abstract}
We report an anomalous electronic transition at 140\text{K} in high-quality Bi$_2$Se$_3$, where charge order emerges in a defect-tuned system. Native defects (Se vacancies and Bi intercalation)-intrinsic to our reproducible growth method-modulate electronic states without compromising sample integrity, mirroring doping-induced phases in correlated topological materials. The hexagonally deformed Fermi surfaces and strong nesting in Bi$_2$Se$_3$ and related compounds (e.g., Bi$_2$Te$_3$) have long suggested the possibility of density wave ordering, with recent work on superconducting Cu- and Nb-doped Bi$_2$Se$_3$ further highlighting charge order's role in unconventional superconductivity. Here, we identify a periodic lattice distortion near room temperature via electron diffraction, consistent with diffuse charge order. This is accompanied by a 140\text{K} electronic transition, manifested in resistivity measurements as a pronounced anomaly, exhibiting a semiconductor-like upturn, signaling the opening of an energy gap. Nuclear magnetic resonance (NMR) studies of the $^{209}$Bi spin-lattice relaxation rate ($1/T_1$) reveal a concurrent transition, confirming the emergence of an 8\textit{\text{meV}} energy gap. Our results are consistent with defect-stabilized charge order in Bi$_2$Se$_3$, linking native defects to its electronic properties and offering broader insights into the interplay between charge order and superconductivity in topological materials.
\end{abstract}


\end{frontmatter}


\section{Introduction}

The 2D layered chalcogenide Bi\textsubscript{2}Se\textsubscript{3} and its intercalated variants have emerged as an ideal system for investigating the interplay between topological order and correlated electronic states.\cite{1,2,3,4,5,6}. As a prototypical Z\textsubscript{2} topological insulator (TI), Bi\textsubscript{2}Se\textsubscript{3} hosts robust Dirac surface states (velocity of $5\times10^5$ m/s) protected by time-reversal symmetry, coexisting with a 0.3 eV bulk band gap \cite{7}. Its electronic properties are exceptionally tunable - through either intrinsic defects (Se vacancies, Bi intercalants) \cite{8,9,10,11} or extrinsic doping (Cu, Sr, Nb) - enabling controlled access to diverse collective phenomena including charge density waves (CDW) and unconventional superconductivity (SC) \cite{12,13,14}. While most unconventional superconductors are driven by strong electron-electron correlations \cite{15,16,17,18}, Bi\textsubscript{2}Se\textsubscript{3} is known to exhibit weak \textit{sp} electron correlations \cite{19,20}, raising the question of alternative mechanisms for unconventional pairing. This unique combination of topological protection and defect-tunable correlations makes Bi\textsubscript{2}Se\textsubscript{3} an ideal platform for exploring how quantum geometry governs emergent many-body states in low-dimensional systems.

Bi\textsubscript{2}Se\textsubscript{3} crystallizes in a layered van der Waals structure that facilitates incorporation of native defects (Se vacancies, Bi intercalants) and foreign dopants (Cu, Sr, Nb) during high-temperature growth. Intercalants occupy van der Waals gaps, enabling charge transfer and anisotropic lattice distortion \cite{1,2,20}. Rapid thermal quenching stabilizes these defects, producing observable SAED modulations including superlattice reflections and stacking faults that break crystal symmetry \cite{21,22,23}. These modifications introduce excess carriers into van der Waals gaps, altering both the local electronic environment and lattice structure. Such anisotropic distortions serve as critical precursors for CDW formation\cite{21,24,25,26,27,28}.

The Fermi surface topology of Bi\textsubscript{2}Se\textsubscript{3} is highly sensitive to both native defects and chemical doping. In pristine Bi\textsubscript{2}Se\textsubscript{3}, ARPES reveals a Dirac cone that evolves from circular to hexagonal (above 350 meV) to hexagram-like (near 435 meV), with flat hexagonal regions generating nesting vectors $\mathbf{Q}_i=2k_F\mathbf{e}_i$ conducive to CDW formation \cite{29,30}. Intercalation induces distinct reconstructions: Cu\textsubscript{x}Bi\textsubscript{2}Se\textsubscript{3} and Sr\textsubscript{0.1}Bi\textsubscript{2}Se\textsubscript{3} exhibit long-wavelength Fermi surface nesting leading to A\textsubscript{2u} superconductivity \cite{1,2}, while experimental studies suggest Nb doping could promote CDW states through multiple Fermi surfaces and local structural distortions \cite{3}. However, direct experimental evidence for CDW formation in doped Bi\textsubscript{2}Se\textsubscript{3} remains elusive, highlighting the need for further investigation of this predicted instability.

Similar CDW transitions are observed in layered chalcogenides ( NbSe\textsubscript{2}, TaS\textsubscript{2}, Cu\textsubscript{x}TiSe\textsubscript{2}), where periodic lattice distortions (PLD) and partial/full Fermi surface gap openings accompany the transition\cite{25,27,28,31,32}. These systems exhibit commensurate or incommensurate modulations in both lattice and electronic structure, typically causing semiconductor-like resistivity increases without full metal-insulator transitions. While Bi\textsubscript{2}Se\textsubscript{3} is a 3D bulk material, its significantly weaker interlayer coupling compared to NbSe\textsubscript{2} \cite{33} suggests that any emergent CDW-like state may exhibit reduced dimensionality (1D or 2D), as observed in the anisotropic NMR, transport, and magnetic measurements of related topological chalcogenides \cite{34,35}. This shared phenomenology underscores the universal role of Fermi surface nesting and electron-phonon coupling in driving collective phases across layered materials, while highlighting dimensionality as a key tuning parameter \cite{24,25}.

In this study, we employ a controlled crystal growth protocol involving rapid thermal quenching to preserve the intrinsic defect configurations, enabling systematic investigation of CDW-like behavior in both native defect-tuned pristine Bi\textsubscript{2}Se\textsubscript{3} and Nb-doped Bi\textsubscript{2}Se\textsubscript{3} (to be reported separately). All samples consistently show CDW-like behaviors: (1) room-temperature SAED patterns reveal superlattice reflections and diffuse scattering, indicating periodic lattice modulations and disorder-perturbed fluctuation; (2) a well-defined resistivity anomaly centered at 140~K; (3) \textsuperscript{209}Bi NMR spin-lattice relaxation rate ($1/T_1$) shows a coherence peak at the transition, followed by thermally activated behavior, signaling the formation of a partial gap; and (4) additional phonon softening at 200~K, suggesting strong electron-phonon coupling preceding the transition. The gradual resistivity feature---combined with the well-defined NMR gap and structural modulations---suggests a spatially inhomogeneous transition, consistent with defect-mediated CDW formation. This reproducible phenomenology establishes Bi\textsubscript{2}Se\textsubscript{3} as a model system where native defects reliably generate electronic instabilities, offering new opportunities for defect engineering in topological materials.

\section{Materials and Method}

Single crystals of Bi\textsubscript{2}Se\textsubscript{3} were prepared by melting high-purity (99.999\%) powders of Bi and Se. Stoichiometric mixtures of 2.5 g were sealed into high-quality quartz tubes in vacuum after being weighed and sealed in an inert glove box, taking extreme care never to expose to air. The mixtures sealed in quartz tubes were heated up to (\( 850^\circ \text{C} \)) and held for 20 hours. They were then cooled to (\( 650^\circ \text{C} \)) at (\( 0.1^\circ \text{C} \))/min, followed by quenching from high temperature into ice water. This yielded large, shiny single crystals which were easily cleaved along the ab plane.

Powder X-ray diffraction data were collected from pieces of single crystals powdered inside an inert glove box. Rietveld refinement was performed using GSAS (General Structure Analysis System) and the EXPGUI interface. Selected Area Electron Diffraction (SAED) was performed at room temperature with a Hitachi H-9000NAR high-resolution transmission electron microscope (HRTEM) operated at 300 \textit{kV}. Four-probe resistivity measurements were performed with a rate of 1K/step at varying temperatures and magnetic fields using a Quantum Design Physical Property Measurement System (PPMS).

Pulsed \textsuperscript{209}Bi NMR (Nuclear Magnetic Resonance) measurements were performed on a Bi\textsubscript{2}Se\textsubscript{3} single crystal of crystal size ~0.94 x 0.58 x 0.41 cm placed inside a home-built probe in an 11-Tesla Helium cryostat. The single crystal of Bi\textsubscript{2}Se\textsubscript{3} was studied with the magnetic field oriented in two directions, \( \mathbf{H} \perp\mathbf{c} \)-axis and \( \mathbf{H} \parallel \mathbf{c} \)-axis. Field sweep spectrums were performed at frequency f=67.875MHz. Spin-echo signals for \textsuperscript{209}Bi NMR spectra were processed using the summed Fourier transform method, with the field swept from 9.7T to 10.5T. Spin Lattice relaxation time $T_1$ measurements in both \( \mathbf{H} \parallel \mathbf{c} \)-axis and \( \mathbf{H} \perp\mathbf{c} \)-axis (H=9.86T) directions were performed at stabilized temperature points varying between 1.6K and 300K. We employed a train of RF pulses on the central transition to saturate the magnetization followed by variable delays and integrate the spin-echoes to map the magnetization recovery. 

The magnetization recovery M(t) is fitted with a stretched single exponential, appropriate for the initial condition where the quadrupolar satellites are completely saturated \cite{36}:
\begin{align}
 M(t) &= M_{\infty}[1- e^{-(t/T_{1})^\beta} ] 
\end{align}
where the parameter $\beta$ allows for a distribution of $T_1$. Magnetization recovery M(t) can either be fitted by a master equation or a stretched exponential equation \cite{36,37}. Stretched exponential fitting is used when there is a continuous distribution of relaxation rates. We used a pulse train to saturate all the NMR lines in our measurement set-up. Although the signal does not saturate completely, we find that a stretched single exponential fits our data better than the master equation.

\section{Results and Discussion}

\subsection{Structural study}

\begin{figure}[ht]
  \centering
   \includegraphics[width=0.8\textwidth] {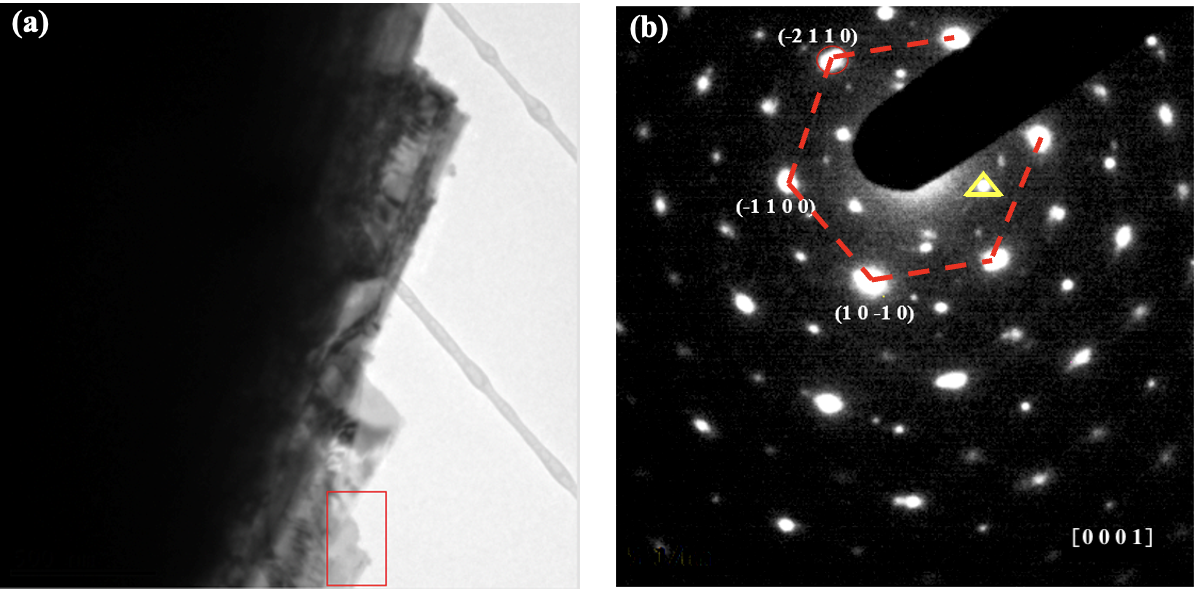}
   \caption{Transmission Electron Microscopy (TEM) and Selected Area Electron Diffraction (SAED) of a Bi\textsubscript{2}Se\textsubscript{3} single crystal at room temperature. (\textbf{a}) Bright-field TEM image of a crystal flake. The red box indicates the region used for SAED. (\textbf{b}) SAED pattern along the [0001] zone axis. The strong reflections (connected by red dashed lines) index to the pristine rhombohedral lattice. The weak, forbidden reflections (indicated by yellow triangles) are commonly associated with stacking faults or other defects in the van der Waals gaps.}
   \label{fig1 }
\end{figure}

\subsubsection{Local structure reveals stacking faults and a periodic distortion}
Figure 1 and 2 show transmission electron microscopy (TEM) bright-field images of Bi$_2$Se$_3$ flakes and their corresponding selected area electron diffraction (SAED) patterns. As shown in Fig. 1(a) and Fig. 2(a), the samples are oriented along the [0001] zone axis and exhibit a layered morphology. The SAED pattern in Fig. 1(b), taken from the area highlighted in Fig. 1(a), shows the primary diffraction spots of the rhombohedral Bi$_2$Se$_3$ phase alongside additional, weaker spots at the $2/3$ reciprocal lattice spots. These are kinematically forbidden reflections that indicate a breakdown of the standard ABC stacking sequence \cite{21, 22, 24, 25, 36, 37, 38, 39}. To probe the origin of these features, the crystal (Fig. 2(a), 2(c)) was tilted slightly off the [0001] zone axis. The resulting SAED patterns (Fig. 2(b), 2(d)) reveal diffuse scattering streaks between the primary Bragg reflections. The presence of these streaks only upon tilting, and their directionality, indicates scattering that is sharp in the $(h k i 0)$ plane but highly diffuse along the $[000l]$ direction, characteristic of disorder primarily along the $c$-axis.The persistence of both features under tilting confirms their structural origin. The streaks are a direct signature of a periodic lattice distortion (PLD) \cite{23, 24, 25, 27, 31, 32}.


\begin{figure}[ht]
  \centering
   \includegraphics[width=0.65\textwidth] {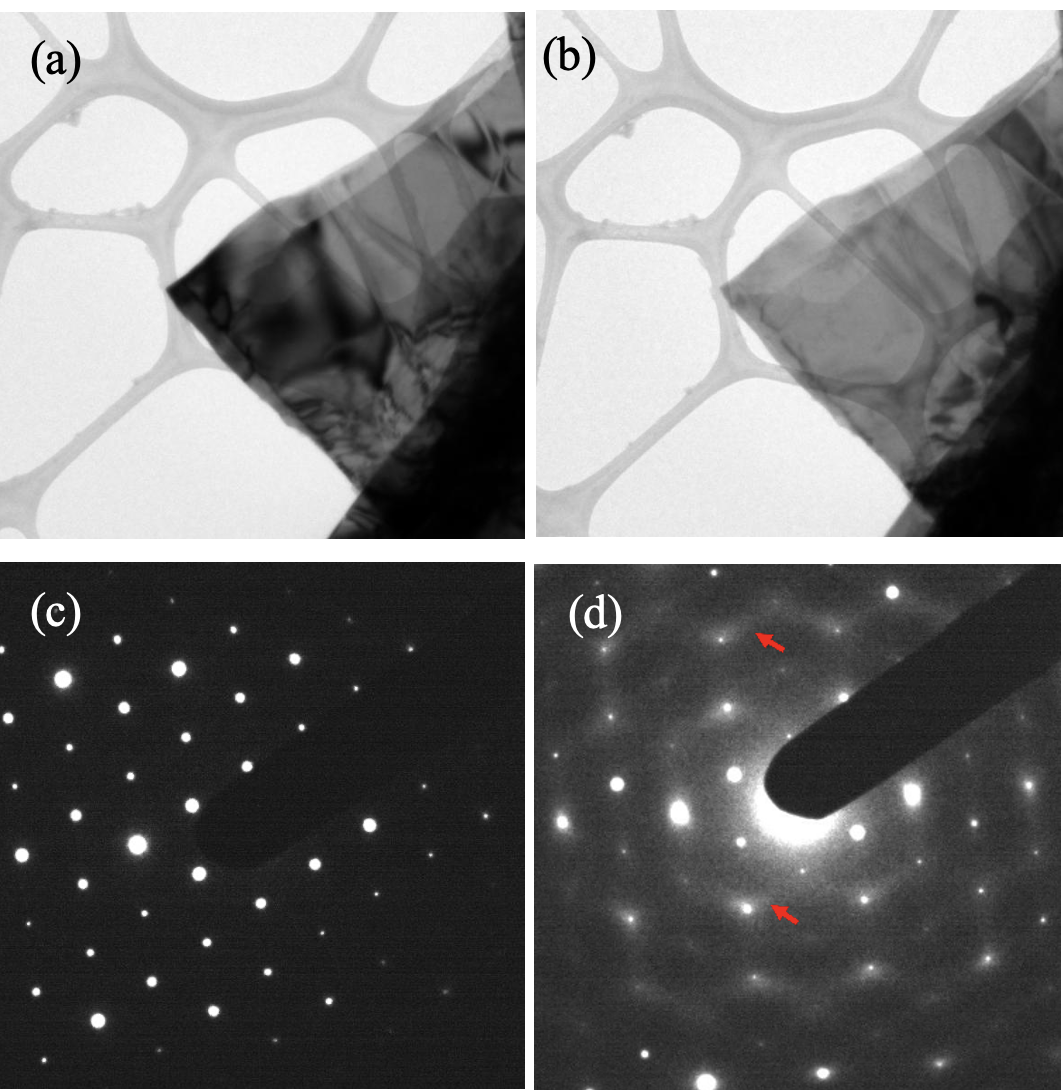}
   \caption{(a), (b) Bright field TEM on a flake obtained from single crystal Bi\textsubscript{2}Se\textsubscript{3}. (c), (d) Selected Area Electron Diffraction from the corresponding areas shown in (a) and (b). The images in (a) and (c) show results when the beam is on [0001] axis. Images in (b) and (d) show off-axis electron diffraction, for which the sample was tilted slightly (less than 5 degrees) away from the zone axis, arrows indicating examples of the diffused regions. }
   \label{fig2 }
\end{figure}

\subsubsection{Powder XRD analysis indicates point defects} 
To understand the driving force for these local modifications, we turned to powder X-ray diffraction (XRD). Rietveld refinement confirms the in-plane lattice parameters $a = b = 4.14$ \AA, but reveals an elongated $c$-axis lattice parameter of $28.66$ \AA. Williamson-Hall analysis of the XRD peak broadening indicates the presence of isotropic lattice strain (see supplemental materials). Both the $c$-axis expansion and the isotropic strain are hallmark signatures of point defects, such as Bi self-intercalation or Se vacancies \cite{40}.

\subsubsection{Synthesis of evidence} The combined SAED and XRD results provide a consistent picture: the point defects detected in the crystal (likely intercalation or vacancies) are the probable cause of the local crystallographic modifications—the stacking faults and the concomitant PLD. In related materials, such a PLD is a known precursor to a charge density wave (CDW) state at lower temperatures \cite{24}. The consequences of these defects for electronic properties are explored below.



\subsection{Transport Measurements}
\begin{figure}[ht]
  \centering
   \includegraphics[width=\textwidth] {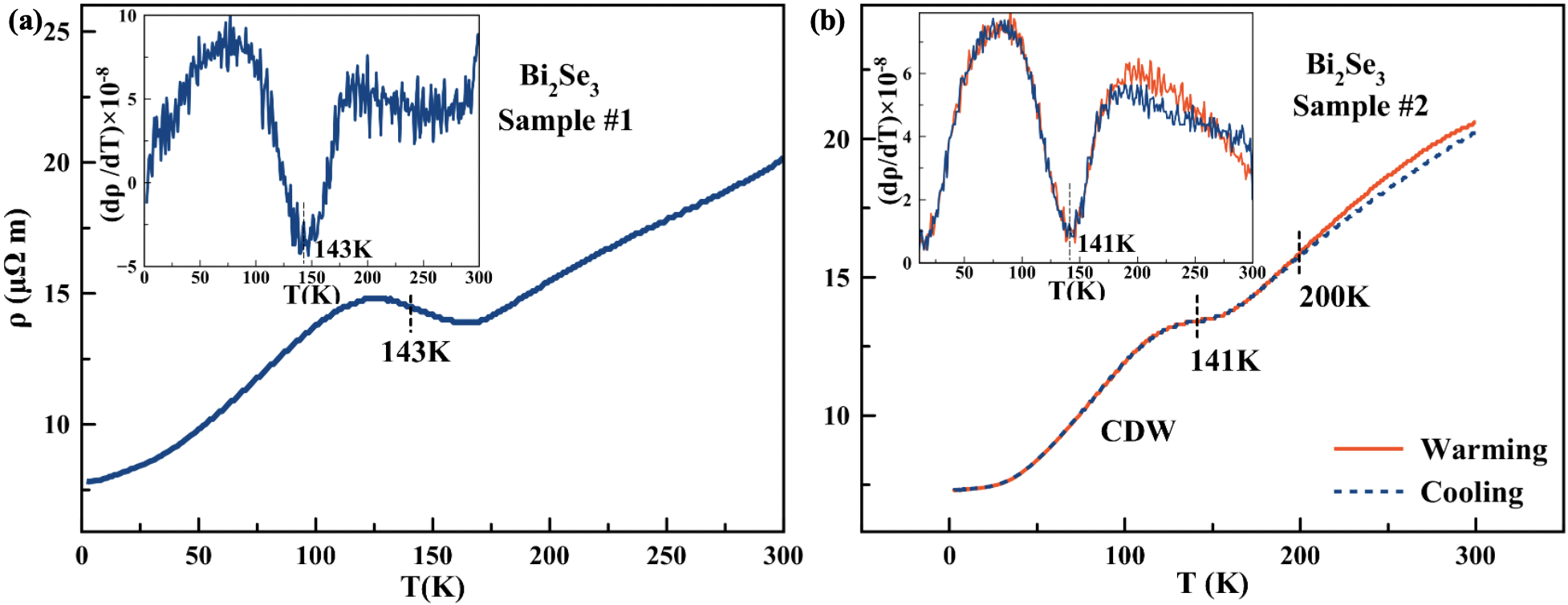}
   \caption{\small Resistivity of Bi\textsubscript{2}Se\textsubscript{3} as a function of temperature for two different pieces, sample $\#$1 and sample $\#$2, measured in zero magnetic field with the electric field along the ab plane. (a) Measurement on sample $\#$1 while cooling the sample from 300 K to 2 K; (b) Measurement on sample $\#$2 for both cooling and heating. Insets for (a) and (b) are the plots of (\textit{d$\rho$/dT}) as a function of temperature to clarify the temperature values of the inflection points and the potential CDW transition temperature \( T_{CDW} \). }
   \label{fig:resistivity }

\end{figure}

\begin{figure}[ht]
  \centering
   \includegraphics[width=0.7\textwidth] {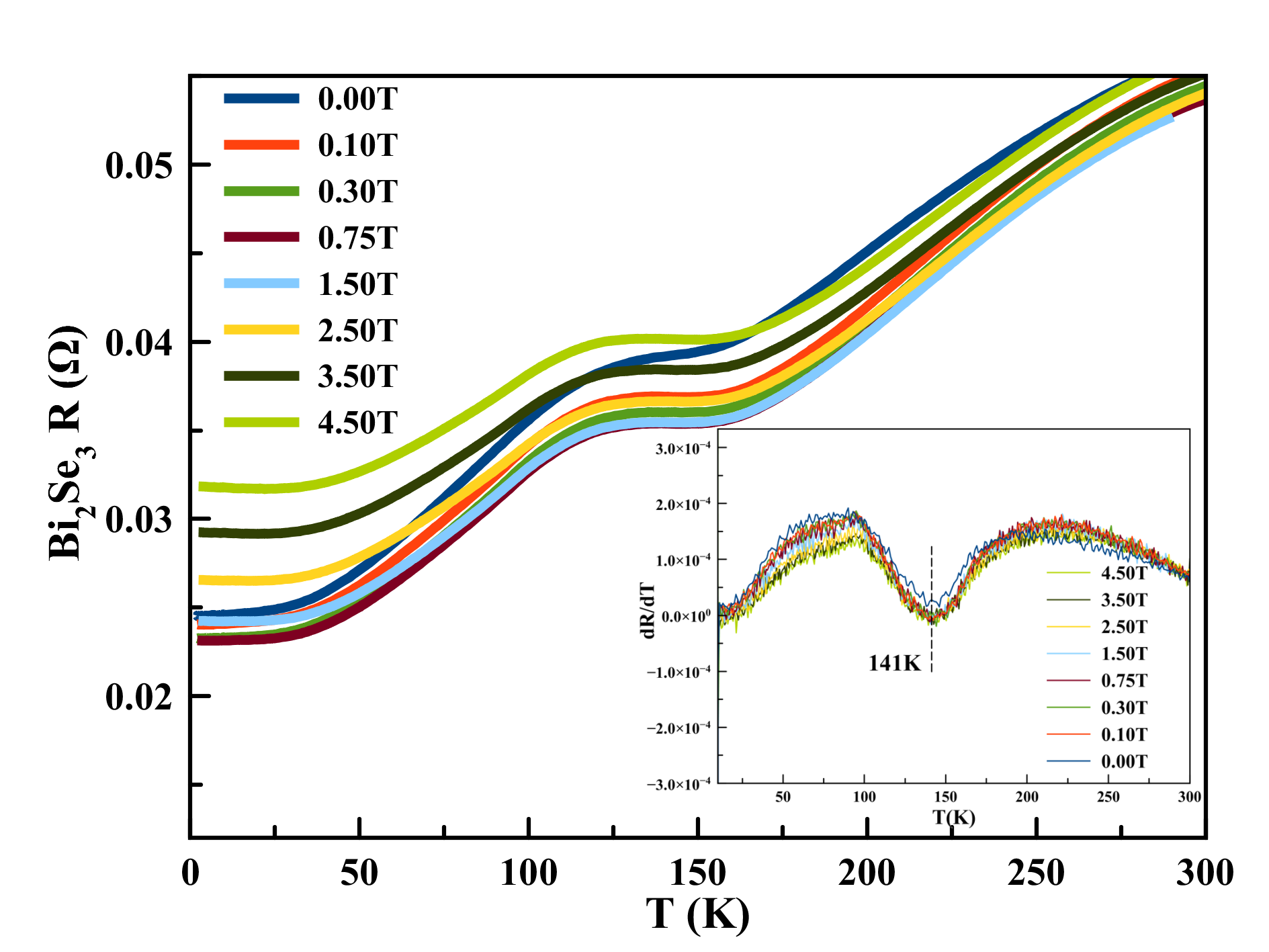}
   \caption{\small Thermal dependence of resistivity of Bi\textsubscript{2}Se\textsubscript{3} for different values of applied magnetic field \( \mathbf{H} \parallel \mathbf{c} \)-axis-axis varying between 0.00 – 4.50 Tesla. }
   \label{fig:resistivity }
\end{figure}

\subsubsection{Anomalous Resistivity and Possible CDW Transition in Bi\textsubscript{2}Se\textsubscript{3}}
Four-probe DC resistivity measurements, shown in Figure 3 and Figure 4, were performed in the 2-300K temperature range with linearly aligned electrodes on the ab surface of Bi\textsubscript{2}Se\textsubscript{3} single crystals, and with the electric field \( \mathbf{E} \parallel \mathbf{ab} \). To ensure reproducibility, we performed measurements on several different pieces of as-grown single crystal. Figures 3(a), 3(b) are the results of different pieces from the same batch of as-grown single crystal. Resistivity measurements at zero ﬁeld all show approximately metallic behavior from room temperature down to around 140K. Note a sharp upturn in resistivity with decreasing temperature, centered around 140K, followed by a return to metal-like behavior below 140K. In Figure 3(a), resistivity rises from ~1.38 × 10\textsuperscript{-5}$\Omega$m to 1.47 × 10\textsuperscript{-5}$\Omega$m upon cooling, while in Figure 3(b), it increases from ~1.31 × 10\textsuperscript{-5}$\Omega$m to 1.36 × 10\textsuperscript{-5}$\Omega$m, both of which are followed by a return to metallic behavior at lower temperatures. The insets in Figures 3(a) and 3(b) show the temperature derivative of resistivity (\textit{d$\rho$/dT}), which highlights the inflection point and onset around 140K. We attribute this metal-to-semiconductor-like behavior to a partial gap or instability at the Fermi level, consistent with a transition into a possible charge density wave (CDW) ground state \cite{44,45}.


\subsubsection{Minimal Hysteresis Confirms Second-Order Nature at 140 K}
Heating and cooling cycles on a second sample (Figure~3b) were performed to assess thermal hysteresis. The key finding is minimal hysteresis around the 140~K transition, which is consistent with a second-order charge density wave (CDW) transition~\cite{46,47}. This observation supports a phonon-mediated mechanism for this transition~\cite{1,2,49,50}.
Separately, we note a subtle hysteresis loop emerges at a higher temperature of $\sim$200~K (Figure~3b). The origin of this weaker anomaly is distinct from the primary 140~K transition and remains uncertain. Without corroborating thermodynamic signatures, it may be attributed to a crossover effect or phonon softening, which could induce sluggish lattice dynamics near this temperature.

\subsubsection{Robustness of the 140 K Transition Under Magnetic Field}
Figure 4 illustrates the magnetic field dependence of the \( 140 \, \text{K} \) transition for sample $\#$2, the same sample shown in Figure 1(b), with the magnetic field (\( \mathbf{H} \parallel \mathbf{c} \)-axis) varying between 0.00 T and 4.50 T. The inset of Figure 4 shows \textit{d$\rho$/dT} as a function of temperature (T), revealing that the transition temperature remains unchanged under magnetic fields below 5 T. This rules out magnetic-field-induced localization effects but closely resembles the unconventional CDW reported in La\textsubscript{3}Co\textsubscript{4}Sn\textsubscript{13}~\cite{46}. The consistency with NMR results (discussed later) further supports this interpretation.

\subsection{Evidence of 140 K and 200 K anomalies in \textsuperscript{209}Bi NMR Spin-Lattice Relaxation}

\begin{figure}[htbp]
  \centering
   \includegraphics[width=\textwidth] {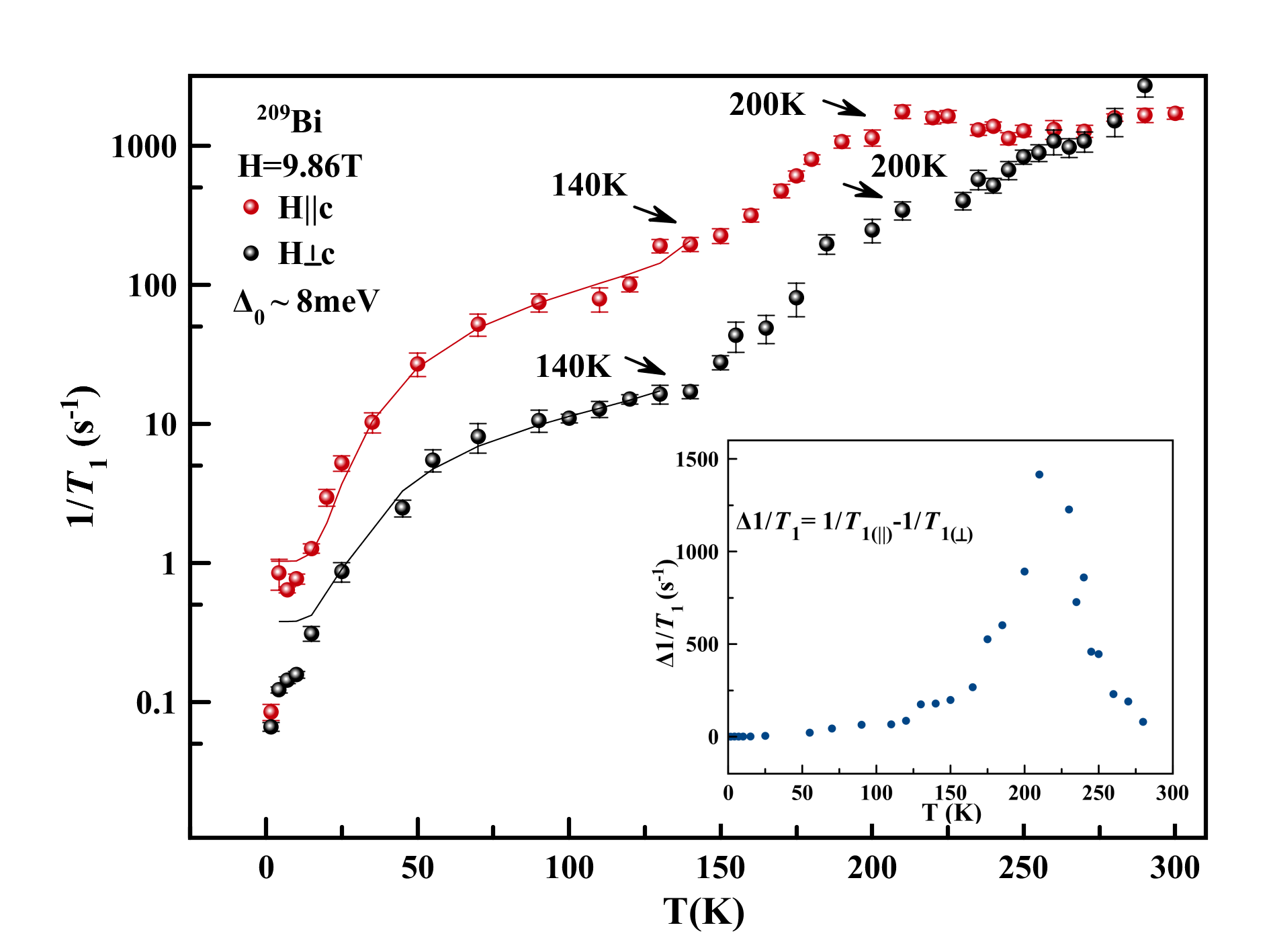}
   \caption{\small Spin-lattice relaxation rate (1/T$_1$) shown with temperature T varying between 1.6K and 300K with applied magnetic field directions: \( \mathbf{H} \parallel \mathbf{c} \)-axis (red balls) and \( \mathbf{H} \parallel \mathbf{c} \)-axis (black balls). Arrows indicate possible transitions at \( 140 \, \text{K} \), and \( 200 \, \text{K} \) in each direction, where \( 140 \, \text{K} \) is the possible commensurate CDW transition temperature. Solid lines are fits to a temperature-dependent CDW energy gap $\Delta_0$, with gap value of $\sim$8 \text{meV}. 
\textbf{Inset:} Differences of 1/T$_1$ from the two magnetic field orientations as a function of temperature, where $\Delta$1/T$_1$ = 1/T$_1$($\parallel$) - 1/T$_1$($\perp$). We see the largest difference occurring at \( 200 \, \text{K} \). }
   \label{fig:resistivity }

\end{figure}

To gain microscopic insight into the nature of these transitions, we performed NMR relaxation measurements. The nuclear spin-lattice relaxation rate, \(( 1/{T_1} \)), quantifies the recovery of the longitudinal nuclear magnetization following an external perturbation, such as an radiofrequency field. The magnetization recovery, M(t), was fitted using a stretched single exponential (see methods). However, we found that the results for $T_1$ were similar whether or not the stretch parameter was included, with the stretched component $\beta$ values close to 1. The results of this fit are shown in Figure 5. Figure 5 presents the temperature dependence of \(( 1/{T_1} \)) for the central $\langle \pm \frac{1}{2} | \leftrightarrow | \mp \frac{1}{2} \rangle$ transition of the \textsuperscript{209}Bi nuclei (I = 9/2), with temperature ranging from 1.6 K to 300 K, pulse separation time $\tau$ = 10 $\mu$ \text{s}, and an external magnetic field of H = 9.86 T (frequency f = 67.875 \, \text{MHz}) applied in two orientations: \( \mathbf{H} \parallel \mathbf{c} \)-axis and \( \mathbf{H} \perp \mathbf{c} \)-axis. It is evident that the spin-lattice relaxation rate, \(( 1/{T_1} \)), exhibits similar temperature-dependent behavior for both field orientations. While the relaxation rates for \( \mathbf{H} \parallel \mathbf{c} \) are roughly an order of magnitude longer compared to \( \mathbf{H} \perp \mathbf{c} \), they converge to nearly the same value at room temperature.



Figure 5 shows two distinct slope changes at \( 140 \, \text{K} \) and \( 200 \, \text{K} \), indicated by arrows. The \( 140 \, \text{K} \) anomaly coincides with the anomaly observed in our resistivity measurements, which we previously interpreted as a transition to a possible commensurate CDW state. The sharp decrease in \(( 1/{T_1} \)) below \( 140 \, \text{K} \) suggests the opening of an energy gap associated with the formation of a CDW-like state below this temperature \cite{51,52,53,54}. The \( T_1 \) data are then fitted to a BCS-like gap equation, as shown in Figure 5. The resulting CDW gap value, approximately 8 \textit{\text{meV}}, is isotropic, as it is independent of the magnetic field direction. The anomaly at \( 200 \, \text{K} \) shows a weak anisotropy with respect to the applied magnetic field. In the inset of Figure 5, we present the anisotropy of the relaxation rate, \( \Delta 1/{T_1} \), as a function of temperature. This anisotropy is small at room temperature and increases as the temperature decreases, peaking at \( 200 \, \text{K} \) before decreasing again towards zero at the base temperature. A more detailed discussion of the \( 140 \, \text{K} \) gap and the \( 200 \, \text{K} \) anomaly is provided in the following section.

\subsubsection{Possible CDW fluctuations at \( 140 \, \text{K} \) analyzed by temperature-dependent BCS gap model }

Nuclear relaxation measurements (Figure 5), along with resistivity measurements (Figures 3 and 4), reveal a novel anomaly near \( 140 \, \text{K} \). We interpret this anomaly as a transition to a possible charge density wave (CDW) state below \( 140 \, \text{K} \). This conclusion is further supported by electron diffraction measurements at room temperature (Figures 1 and 2), which show evidence of a periodic lattice distortion (PLD) combined with diffuse scattering in off-axis diffraction, suggesting the presence of a CDW instability that develops into a possible charge density wave state at lower temperature. \( T_1 \) anomalies have been previously linked to CDW-like transitions in layered systems \cite{51,52,53,54}.

Below \( 140 \, \text{K} \), the CDW gap can be determined from the relaxation rate, as follows \cite{56,57}:
\[
\frac{1}{T_1} = a \exp\left(-\frac{\Delta(T)}{T}\right) + c \tag{2}
\]
where we use the form of a temperature-dependent BCS gap equation:
\[
\Delta(T) = \Delta_0 \tanh\left(1.74 \sqrt{\frac{T_C}{T} - 1}\right) \tag{3}
\]
Here, \( T_C = T_{CDW} \) is the CDW transition temperature, and \( a \) and \( c \) are fitting constants. The fit (Figure 5) yields a energy gap of \( \Delta_0 = 8 \pm 2 \, \text{\text{meV}} \) for both directions of the applied field. We estimate the coupling constant \( \lambda = \frac{\Delta_0}{k_B T_{CDW}} = 0.7 \) (where \( T_{CDW} = 140 \, \text{K} \) and \( k_B \) is the Boltzmann constant), which falls within the weak-coupling regime of BCS theory (\( \lambda \sim 0.5-1.2 \)) \cite{58} but is smaller than typical values for conventional phonon-mediated superconductors (where \( \lambda \sim 1.76 \)). This suggests that additional mechanisms beyond standard electron-phonon coupling may be at play. Several factors could contribute to this reduced coupling strength:
\begin{itemize}
\item{Weaker electron-phonon interaction in the CDW state}: the smaller energy gap (\( \Delta_0 = 8 \pm 2 \, \text{\text{meV}} \)) compared to BCS expectations implies a weaker coupling strength, possibly due to a reduced electronic density of states or anisotropic gap formation near the Fermi level.
\item{Dimensionality effects}: if the CDW state has 1D or 2D character, enhanced fluctuations or reduced electronic screening-compared to 3D systems-could suppress the effective coupling constant . 
\item{Competition with other electronic orders}: In Bi\textsubscript{2}Se\textsubscript{3}, the coexistence of superconductivity and topological surface states may compete with CDW formation \cite{12,13,14}, renormalizing the electron-phonon interaction and leading to a weaker \(\lambda\).
\end{itemize}

\subsubsection{Comparisons of our results with previous $^{209}$Bi NMR and transport work in Bi\textsubscript{2}Se\textsubscript{3}}

\begin{figure}[htbp]
  \centering
   \includegraphics[width=\textwidth] {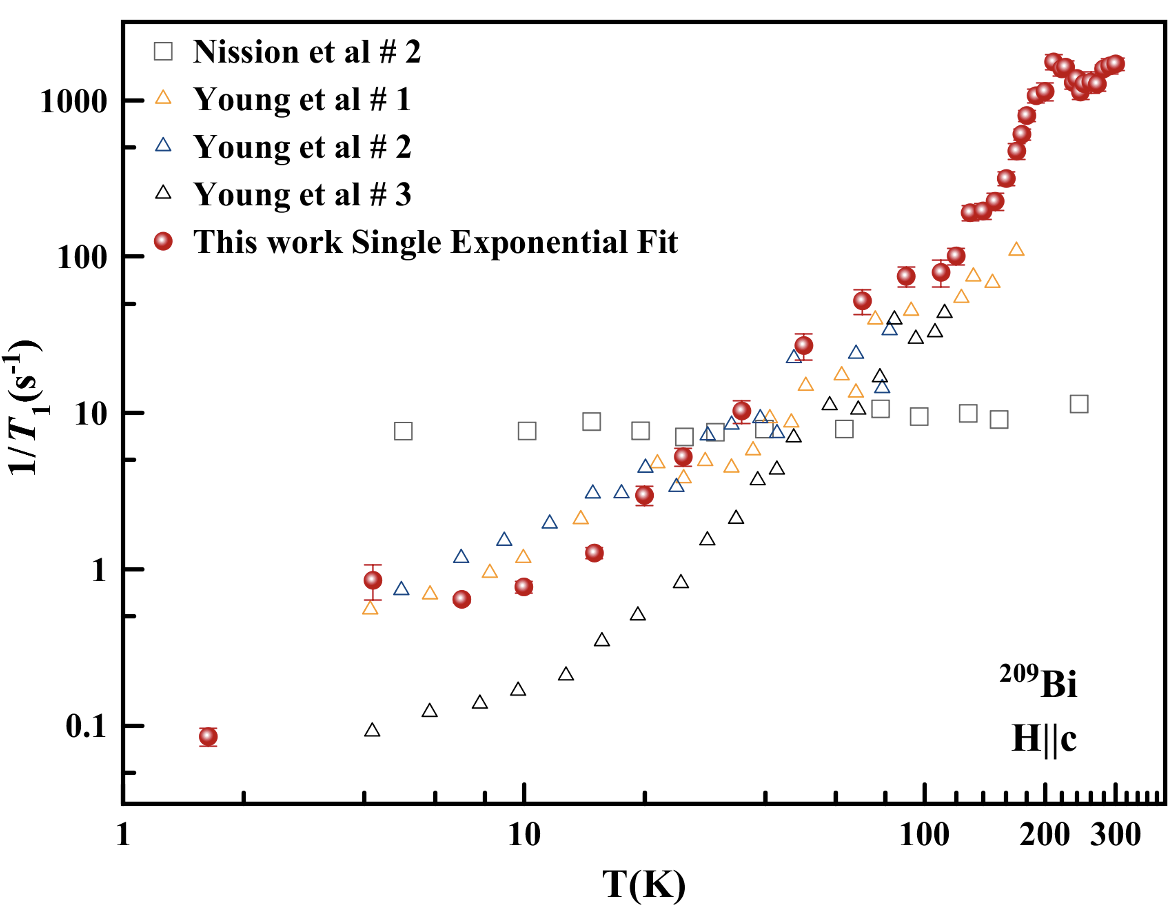}
   \caption{\small Comparison of Spin-lattice relaxation rates of $^{209}$Bi versus temperature from earlier published papers to our results in the magnetic field \( \mathbf{H} \parallel \mathbf{c} \)  direction. The open data points are reproduced from Ref. \cite{59} (square shapes from Nisson’s sample $\#2$, most of Nisson’s results are in the same range) and Ref. \cite{60} (triangle shapes, sample $\#1$, $\#2$, and $\#3$). Solid data points are our results, the same as displayed in Figure 5 \( \mathbf{H} \parallel \mathbf{c} \)  direction data, with a single exponential fit. }
   \label{fig:resistivity }
\end{figure}
Previous relaxation measurements have shown significant variation in magnitudes and temperature dependence across samples prepared using different methods \cite{59,60}. These variations were partly attributed to defects associated with Se vacancies, which could shift the Fermi level by partially occupying the conduction band. In contrast to previous studies, our NMR \( T_1 \) data show a very strong temperature dependence from 1.6 to 300 K, spanning nearly four orders of magnitude for \( \mathbf{H} \parallel \mathbf{c} \)-axis  (see Figure 6 for all \( T_1 \) data). The low-temperature behavior (below 50 K) is similar to that of sample $\#$2 from Ref.\cite{60}, which the authors identify as the sample with a higher carrier concentration.

\begin{figure}[htbp]
  \centering
   \includegraphics[width=\textwidth] {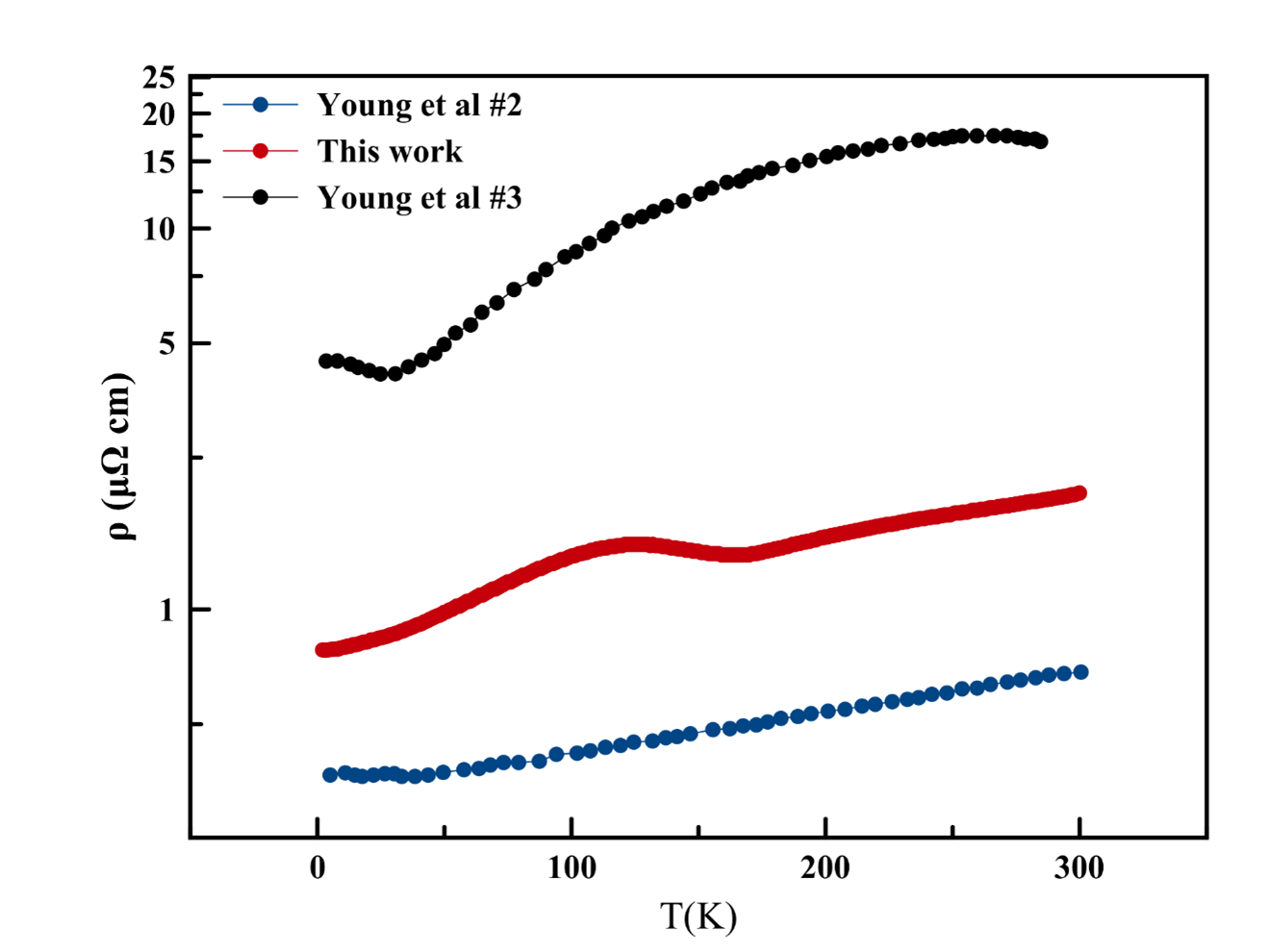}
   \caption{\small Comparison of resistivity vs. temperature results from our transport measurement (red) and Ref. \cite{60} Young et al. (sample $\#2$ blue and sample $\#3$ black). In the plot, the units are unified to ``$\mu \Omega$ cm''.}
   \label{fig:resistivity }
\end{figure}

Further, a comparison of the resistivity data (Figure 7 — composite plot of $\rho$ from Ref. \cite{60} Young et al. and this work) reveals that our sample exhibits slightly fewer carriers than Young’s sample $\#$2, but it remains far from insulating, suggesting that the Fermi level is situated lower in the conduction band. This observation aligns with our NMR Knight shift measurements, which fall between those of Young’s samples $\#$2 and \#3. Specifically, our isotropic Knight shift (K\textsubscript{iso}) is approximately 0.4$\%$, while Young’s sample $\#$2 shows around 0.65$\%$, and sample $\#$3 displays 0.34$\%$ at 4.2 K. One could argue that it is possible for our sample to have serendipitously reached a half-band-filling condition, which is crucial for a Peierls transition, due to doping via Se-vacancies. This may have led to the formation of a charge-density wave (CDW) in our sample, a phenomenon not observed in the others.

\subsubsection{Distinct High-Temperature Anomaly at \( 200 \, \text{K} \) of potential phonon softening}

The NMR anomaly near \( 200 \, \text{K} \) merits further discussion in light of the fact that the Debye temperature is \( \Theta_D \approx 182 \, \text{K} \) for Bi\(_2\)Se\(_3\) \cite{61}. Given that the behavior of \(( 1/{T_1} \)) depends on the details of the high-temperature phonon spectrum, one would expect it to behave differently above and below the Debye temperature. Thus, it is possible that the transition observed around \( 200 \, \text{K} \) in Figure 5 is due to changes in the phonon spectrum across \( \Theta_D \). However, note the anisotropy in this transition with respect to the direction of the applied magnetic field, \( \mathbf{H} \parallel \mathbf{c} \) and \( \mathbf{H} \perp \mathbf{c} \), as is clear from Figure 5. 

In Figure 5, the $(1/T_1)$ data for both orientations exhibit similar temperature dependence, although the relaxation rates for \( \mathbf{H} \parallel \mathbf{c} \) are approximately an order of magnitude faster than those for \( \mathbf{H} \perp \mathbf{c} \). Notably, these values converge to nearly identical values at room temperature. Above \( 200 \, \text{K} \), the temperature behavior differs: the \(( 1/{T_1} \)) value remains constant from room temperature to \( 200 \, \text{K} \) for the parallel case, while it precipitously drops for the perpendicular case. This would imply a constant relaxation mechanism that only exists in the parallel direction and is weakly dependent on temperature, while in the perpendicular direction, this mechanism monotonically diminishes until the temperature reaches \( 200 \, \text{K} \). The origin of this mechanism is not clear, but it may be related to the topological nature of this family of systems, which suppresses the relaxation channel when the field is applied parallel to the plane of the layers \cite{61, 62}.

To analyze this further, let us suppose that the relaxation rates can be separated into electronic and phonon contributions \cite{60}:
\[
\frac{1}{T_1}(k) = \frac{1}{T_1^{\text{el}}}(k) + \frac{1}{T_1^{\text{ph}}}(k) \tag{4}
\]
where \( k = \parallel, \perp \) are the magnetic field directions with respect to the crystal c-axis. The electronic contribution is magnetic in nature and arises from the spin scattering of the conduction electrons. On the other hand, the phonon contribution is driven by changes in the lattice, which influence nuclear relaxation via quadrupolar effects. Since the CDW energy gap we obtained below \( 140 \, \text{K} \) is isotropic, it is reasonable to assume that the conduction band electronic contribution \( \frac{1}{T_1^{\text{el}}}(k) \) is also isotropic, as well as the temperature dependence of \( \frac{1}{T_1^{\text{el}}}(k) \) itself. 

We plot in the inset of Figure 5 the differences in relaxation rates \( \Delta \frac{1}{T_1} = \frac{1}{T_1}(\parallel) - \frac{1}{T_1}(\perp) \) between each field orientation. Here, we can clearly see that the anisotropy in relaxation is mainly due to the phonon contribution along the \( \mathbf{H} \parallel \mathbf{c} \) direction, i.e., \( \Delta \frac{1}{T_1} > 0 \) at all temperatures. It starts small at room temperature, peaks at \( 200 \, \text{K} \), and finally freezes as the sample is further cooled.

The positive value of \( \Delta \frac{1}{T_1} \) indicates stronger lattice vibrations (phonons) in the \( \mathbf{H} \parallel \mathbf{c} \)-axis measurements. This is consistent with the 2D layered structure of the material, where Bi-Bi bonds within the ab planes exhibit stronger electron-phonon coupling compared to those along the c-axis \cite{63}. The observed anisotropy aligns with expectations, as the longitudinal relaxation rate \( \frac{1}{T_1}(\parallel) \) is sensitive to fluctuations perpendicular to the magnetic field direction. This NMR anisotropy likely originates from phonon softening, which enhances lattice fluctuations near 200 K.

From \( 200 \, \text{K} \) to \( 140 \, \text{K} \), the electronic contribution becomes dominant, and the phonon vibrations are further weakened. However, phonon vibrations are not entirely absent and still contribute to the relaxation. In this range, some electronic charges are ordered with the underlying lattice, and some are not; thus, there could be a mixture of I-CDW and CDW phases. Since \( 1/{T_1} \) is an integration over the entire Fermi surface, it is possible that in this temperature range the Fermi surface only opened a small segment with mini CDW gaps rather than full Fermi surface reconstruction. At $140\,\text{K}$, a critical threshold is reached: phonon coupling diminishes sufficiently to stabilize stronger charge-lattice commensuration. However, given the persistence of metallic behavior, this transition likely represents a crossover to a \textit{partially gapped}  CDW-like state rather than a complete Peierls transition. The residual conductivity suggests either a nodal gap structure or phase fluctuations that prevent full gap formation. We propose future investigations into the Fermi surface geometry of this material.

\subsection{Possible Mechanisms of the CDW Origin and the Critical Role of Crystal Growth}

The exact nature of Fermi surface distortion, nesting, and density wave instability in systems like Bi\textsubscript{2}Se\textsubscript{3} is critically dependent on the details of its fermiology, which is itself highly sensitive to sample synthesis conditions. Correlated electron systems typically exhibit intertwined electron-ordered states arising from multiple degrees of freedom, including lattice, charge, dimensionality, nematicity, and spin. The manifestation of these states is strongly materials-dependent and is profoundly influenced by crystalline perfection and stoichiometry. For instance, many properties of chalcogenides such as Bi\textsubscript{2}Se\textsubscript{3} stem from the nature of fluctuations or overlaps in the order parameters of these intertwined states. Below, we discuss how the specific material characteristics of Bi\textsubscript{2}Se\textsubscript{3}, which can be tuned via crystal growth, can give rise to multiple ground states, such as that of a charge density wave (CDW).

\subsubsection{The Origins of the Possible CDW Phase Transition in Bi\textsubscript{2}Se\textsubscript{3}}

While charge density waves (CDWs) are typically associated with low-dimensional systems, their formation in more three-dimensional (3D) materials like Bi\(_2\)Se\(_3\) remains less understood \cite{64}. Although Bi\(_2\)Se\(_3\) possesses a 3D electronic structure, its layered van der Waals nature confers a quasi-2D character. This structure is also susceptible to native defects like Se vacancies and Bi self-intercalation in the van der Waals gaps. Such defects can form clusters or chains that generate significant lattice and charge disorder, promoting electronic instability.

These structural details critically influence Fermi surface anisotropy. For instance, Xiangang and Sergey \cite{1} showed that small changes in the position of Bi in the z-direction of doped-Bi\textsubscript{2}Se\textsubscript{3} systems can modify the nesting vector
\[
X(\mathbf{q}) = \sum \delta(\epsilon_k) \delta(\epsilon_{k+\mathbf{q}}),
\]
which is maximized for a wavevector \( \mathbf{q} \) along \( \Gamma_Z \) near the zone center. This promotes strong Fermi surface nesting and electron-phonon coupling (EPC). A [0001] displacement at small \( q_s \) breaks inversion symmetry, lifts degeneracy, and results in a large EPC matrix element. This theoretical picture is supported by broad phonon linewidths at small \( q_s \) observed in both calculations \cite{1} and neutron scattering experiments \cite{2}.

Building on this picture of a q-dependent EPC mechanism, the origin of the CDW instability in Bi\(_2\)Se\(_3\) remains debated, with proposals ranging from exotic orbital or charge order \cite{3} to more conventional EPC. Our observations—including the direct lattice distortion, phonon softening, and the sensitivity to defects—strongly support a dominant EPC mechanism. This interpretation is firmly grounded in the established phenomenology of electronic crystals \cite{65}. However, the lack of definitive long-range structural evidence suggests a scenario where native defects may pin fluctuations and inhibit full long-range order, akin to mechanisms discussed for other systems \cite{66,67}. Within the broader framework for CDWs in low-dimensional systems \cite{68}, the specific growth conditions of our crystals become critical. Specifically, Bi self-intercalation and Se vacancies create localized strain and charge inhomogeneity that enforce energy instabilities in the van der Waals gap, leading to the electronic disorder discussed in the following section.

\subsubsection{Interplay of Growth Conditions with Lattice and Charge Order}
The somewhat elusive nature of superconductivity in single crystals of doped-Bi\(_2\)Se\(_3\), especially the variability of superconducting fraction and \( T_c \) with quenching temperature \cite{38, 39}, is likely due to the high sensitivity of electron interactions to the actual conditions of quenching and growth. The crystals discussed here were grown using a standard self-flux method, with a quenching temperature of \( 650^\circ \text{C} \). The ideal topological insulator, Bi\(_2\)Se\(_3\), has separate bulk and surface states. However, experimentally, intrinsic defects and disorders are commonly found in Bi\(_2\)Se\(_3\) and intercalated/electron-doped Bi\(_2\)Se\(_3\) \cite{3,9,38}. Schneeloch et al. \cite{39} used different growth conditions to grow Cu-doped Bi\(_2\)Se\(_3\) superconductors and reported that high-temperature quenching (above \( 560^\circ \text{C} \)) is essential for superconductivity, especially superconductivity with a high diamagnetic shielding fraction. Other growth conditions either cause no superconductivity (SC) or show a weak diamagnetic shielding fraction. This is also our observation in Cu-Bi\(_2\)Se\(_3\) \cite{38}. Schneeloch et al. suggest that the quenching process helps maintain either a primary intercalated phase or a secondary phase responsible for superconductivity. Huang et al. \cite{40} show that high annealing temperatures (~600\(^\circ\)C) can cause intercalation of Bi in Bi\(_2\)Se\(_3\). Our results in this article indicate that high-temperature quenching (from above \( 650^\circ \text{C} \)) leads to strong electron and lattice order, and interesting electronic ground states such as a charge density wave (CDW) or superconductivity.

X-ray diffraction (XRD) on powdered samples reveals that the \( a \)-axis value of our single crystal Bi\(_2\)Se\(_3\) agrees with those of most other reports, but the \( c \)-axis, at \( 28.66 \, \text{\AA} \), is \( 0.02 \, \text{\AA} \) higher than the \( 28.64 \, \text{\AA} \) reported in most previous studies of Bi\(_2\)Se\(_3\) \cite{40, 64, 69}. Huang et al. \cite{40} assert that a longer \( c \)-axis arises from unintentionally doped Bi-rich flux growth of Bi\(_2\)Se\(_3\), where Bi forms a neutral metal Bi\(_2\) layer intercalated into the van der Waals gap. They also show that crystals with patches of intercalated Bi exhibit high \( c \)-axis values of up to \( 28.65 \, \text{\AA} \), close to the \( c \)-axis value of \( 28.66 \, \text{\AA} \) obtained from our Rietveld refinement. XRD results on our Bi\(_2\)Se\(_3\) single crystals show no signs of the formation of metastable phases of staged \( (\text{Bi}_2)_m (\text{Bi}_2\text{Se}_3)_n \). 

The solidification temperature of Bi\(_2\)Se\(_3\) (\( 705^\circ \text{C} \)) is higher than the melting point of both pure Bi (\( 271.4^\circ \text{C} \)) and pure Se (\( 220^\circ \text{C} \)). Additionally, Se is a vapor above \( 685^\circ \text{C} \), which is below the solidification temperature of Bi\(_2\)Se\(_3\). Consequently, the stoichiometry of Bi\(_2\)Se\(_3\) forming at the liquid-vapor interface can be highly dependent on the vapor pressure of Se at \( 705^\circ \text{C} \). For high annealing temperatures (around \( 600^\circ \text{C} \)), partial decomposition might occur. Huang presumes that this is due to a large number of Se vacancies created in an evacuated environment, resulting in liquid Bi in the flux ending up in the van der Waals gaps rather than incorporating into a Bi-Se quintuple layer containing Se vacancies. 

Our crystals, quenched at \( 650^\circ \text{C} \) (above the high annealing temperature of \( 600^\circ \text{C} \) that Huang claims results in Se vacancies), could lead to Se vacancies and intercalated Bi. We surmise that, as there is not enough excess Bi to form the metastable phase of staged \( (\text{Bi}_2)_m (\text{Bi}_2\text{Se}_3)_n \), excess bismuth in our crystals forms randomly distributed Bi\(_2\) interlayers in the crystal. The resulting Bi-chains could help form a quasi-1D Peierls-type transition or a quasi-2D CDW \cite{40, 64, 70}. In summary, specific growth conditions can drive the observation of a CDW or superconductivity in Bi\(_2\)Se\(_3\). Further work is needed in this direction.

In addition to the discussion of Figure 7, a comparison of our temperature-dependent resistivity data and Knight shift of the central transition with Young’s sample $\#$2 and sample $\#$3 further supports the notion that crystal growth conditions have a significant impact on the electronic behavior of the samples. All three samples (ours, Young’s sample $\#$2, and sample $\#$3) were grown under different conditions. Young's ARPES results show that the Fermi level of their sample $\#$2 lies within the bottom of the conduction band, while that of sample $\#$3 lies in the surface state, far below the conduction band. Since both our resistivity data and Knight shift values are between those of Young’s two samples (but closer to sample $\#$2), it is likely that the Fermi level of our sample sits close to, but lower than, the Fermi level of their sample $\#$2.

\section{Conclusion}

In conclusion, our multi-probe investigation of high-temperature quenched Bi\textsubscript{2}Se\textsubscript{3} reveals a two-stage electronic reorganization. While room-temperature SAED shows diffuse streaks signaling disorder-perturbed charge density wave (CDW) fluctuations, the transitions observed in resistivity and $^{209}$Bi NMR ($1/T_1$) below $140$~K are consistent with the formation of a partially gapped, CDW-like state. The persistence of metallicity, with a small gap ($\Delta \approx 8$~meV), and the lack of definitive long-range structural evidence suggest that native defects may play a pivotal role in stabilizing this state, potentially by pinning fluctuations and inhibiting full long-range order. The precursor dynamics observed in the ($1/T_1$) anisotropy between $\mathbf{H} \parallel \mathbf{c}$ and $\mathbf{H} \perp \mathbf{c}$ near $200$~K further underscore the complex interplay between defects and charge correlations. Overall, our results establish defect-rich Bi\textsubscript{2}Se\textsubscript{3} as a promising platform for exploring the defect-engineering of correlated, metastable electronic phases, dimensional crossover in quasi-2D systems, and the interplay between charge order and strong spin-orbit coupling.

\section*{Funding}
A portion of this work was performed at the National High Magnetic Field Laboratory, which is supported by the National Science Foundation Cooperative Agreement No. DMR-1644779 and the state of Florida. We acknowledge an earlier AFOSR MURI grant to Prasenjit Guptasarma.
\section*{Declaration of competing interest}
The authors declare that they have no known competing financial interests or personal relationships that could have appeared to influence the work reported in this paper. 
\section*{Data availability}
Data is available upon request. 
\section*{Acknowledgments}
We gratefully acknowledge support from Steve Hardcastle of the Advanced Analysis Facility, and Prof. M. Gajdardziska-Josifovska for the use of the Hitachi H-9000NAR high-resolution transmission electron microscope in the HRTEM facility.

\end{document}